SUBJECT AREAS

Physical sciences (Optics and photonics/Nanoscience and technology/Mathematics and computing)

# Chaotic oscillation and random-number generation based on nanoscale optical-energy transfer

Makoto Naruse,[1,*] Song-Ju Kim,[2] Masashi Aono,[3,4] Hirokazu Hori[5] and Motoichi Ohtsu[6]

[1] Photonic Network Research Institute, National Institute of Information and Communications Technology, 4-2-1 Nukui-kita, Koganei, Tokyo 184-8795, Japan

[2] WPI Center for Materials Nanoarchitectonics, National Institute for Materials Science, 1-1 Namiki, Tsukuba, Ibaraki 305-0044, Japan

[3] Earth-Life Science Institute, Tokyo Institute of Technology, 2-12-1 Ookayama, Meguru-ku, Tokyo 152-8550, Japan

[4] PRESTO, Japan Science and Technology Agency, 4-1-8 Honcho, Kawaguchi-shi, Saitama 332-0012, Japan

[5] Interdisciplinary Graduate School of Medicine and Engineering, University of Yamanashi, 4-3-11 Takeda, Kofu, Yamanashi 400-8511, Japan

[6] Department of Electrical Engineering and Information Systems, Graduate School of Engineering, The University of Tokyo, 2-11-16 Yayoi, Bunkyo-ku, Tokyo 113-8656, Japan

* Corresponding author: naruse@nict.go.jp

**By using nanoscale energy-transfer dynamics and density matrix formalism, we demonstrate theoretically and numerically that chaotic oscillation and random-number generation occur in a nanoscale system. The physical system consists of a pair of quantum dots (QDs), with one QD smaller than the other, between which energy transfers via optical near-field interactions. When the system is pumped by continuous-wave radiation and incorporates a timing delay between two energy transfers within the system, it emits optical pulses. We refer to such QD pairs as nano-optical pulsers (NOPs). Irradiating an NOP with external periodic optical pulses causes the oscillating frequency of the NOP to synchronize with the external stimulus. We find that chaotic oscillation occurs in the NOP population when they are connected by an external time delay. Moreover, by evaluating the time-domain signals by statistical-test suites, we confirm that the signals are**

**sufficiently random to qualify the system as a random-number generator (RNG). This study reveals that even relatively simple nanodevices that interact locally with each other through optical energy transfer at scales far below the wavelength of irradiating light can exhibit complex oscillatory dynamics. These findings are significant for applications such as ultrasmall RNGs.**

Modern neuroscience currently considers that the complex interactions between spiking pulses in human brains are at the origin of intelligence[1]. It is clear that humans cannot live without the rhythmic patterns of signals or material flows such as the circadian rhythm[2]. Moreover, even relatively simple biological organisms such as single-celled amoeboid organisms (e.g., *P. polycephalum*) exhibit complex spatiotemporal dynamics including chaotic oscillatory dynamics[3,4]. This intriguing real-world observation raises the following question: What is the ultimate physical architecture in nature that exhibits complex pulsation dynamics?

To address this question, we use theory and numerical analysis to examine intriguing oscillatory dynamics that are based on optical-near-field-mediated energy transfer at a scale far below the wavelength of irradiating light. The insights obtained herein can help us understand the complex oscillatory phenomena observed in micro- and nanoscale engineering devices and natural biological organisms. Moreover, they can help the development of practical applications such as random-number generators[5], which are critical in cryptography[6] and computer simulations[7], as well as in designing "nano intelligence"[8-10].

Energy transfer based on optical near-field interactions between nanoscale materials has been thoroughly studied by fundamental theory[11,12] as well as experiments[13-16]. Generating periodic optical pulses is one of the most important functions of digital systems[17]. To study the generation of optical pulses based on optical near-field processes in the subwavelength regime, Shojiguchi *et al.* theoretically investigated the generation of super-radiance in *N* two-level systems connected by optical near-field interactions[18]. By substantially simplifying Shojiguchi's architecture, Naruse *et al.* theoretically demonstrated optical pulsation in a system composed of two subsystems, each of which involved energy transfer from a smaller to a larger quantum dot (QD). The energy transfer occurs via optical near-field interactions and is driven by a continuous wave (cw) irradiation[19], which results in the emission by the QD system of an optical pulse train. Thus, we refer to the QDs in this context as "nano-optical pulsers" (NOPs).

In many versatile systems in nature and in engineering devices and systems, synchronization and chaos are important phenomena connected to periodic signals[20-22]. For example, injection locking of lasers is of fundamental importance, and stability, instability, and chaos in such systems have been thoroughly studied from basic and practical perspectives[21,23]. In addition, rather than suppressing such chaotic behavior in lasers, the phenomenon can be exploited by applications that secure data communications[24,25]. In associated research, optical random-number generators (RNGs) were intensively investigated[6,26], and chaos generated in quantum-dot microlasers with external feedback was also reported[27].

However, these studies of synchronized, chaotic, and random oscillatory dynamics require far-field optics, which means that the devices and systems are constrained by the diffraction limit of light. This physical restriction means that *macroscale* devices are inherently required. In contrast, the present study focuses on *nanoscale* oscillatory dynamics, which are free from the diffraction limit imposed by optical far fields. By revealing the basic functions made possible by synchronization and chaos in near-field optics, we provide guiding design principles for future devices, systems and methods to evaluate their performance. Note that the pulsation, synchronization, and chaos, as discussed in this paper, are related to *optical pulses* for which the carrying frequencies are fixed, whereas the conventional literature on synchronization and chaos in lasers[21] discusses the oscillation frequency of the radiation itself.

We now give a brief outline of the paper. Reference [19] discusses the combination of two energy transfers by near-field interactions, one of which is delayed with respect to the other. When pumped by cw irradiation, the system emits an optical pulse train. This phenomenon was explained using a density matrix formalism involving six energy levels. In our study, we first further simplify such a pulse-generating mechanism by replacing one of the energy-transfer paths by a delay function. This approach allows us to confirm the emission of a pulsed output. Second, we demonstrate that such an NOP can be synchronized with a periodic external signal. We show that the synchronization bandwidth depends on the intensity of the external stimulus and that the "sensitivity" (defined later) to the external stimulus increases for weaker cw excitation of the NOP. Third, we characterize bifurcations and chaos by combining NOPs with an external timing delay between energy transfers. Finally, by using security-test suites to evaluate the chaotic signals, we checked the randomness inherent in those signals to determine if such devices can be used as an RNG.

Uchida *et al.* experimentally demonstrated an RNG based on semiconductor lasers and achieved 1.7 Gb/s random-number generation[6]. The rate obtained was excellent and devices were developed on the basis of solid and sophisticated principles in the literature on optical communications. However, because these results are based on far-field optics, they suffer from a fundamental difficulty that they cannot be miniaturized beyond the diffraction limit of light[28]. NOPs, however, are based on energy transfer and thus are not diffraction limited. In addition, cw light sources, such as light-emitting diodes and lasers, have now been developed on the basis of principles of near-field optics[29,30], which suggests that optical pulsation and RNGs can be implemented on the basis of nanophotonic principles and technologies.

**Results**

**Nano-Optical Pulser based on Energy Transfer**

Previous work presented a theory of a pulse-generating mechanism in a system of four QDs. This mechanism combines two energy-transfer pathways in which one pathway experiences a timing delay[19]. Here, we first introduce a simpler theory based on a pair of QDs, with one QD smaller than the other.

In the long-wavelength approximation, the electric-field operator is constant in the Hamiltonian, which describes the interaction between an electron and an electric field, because the electric field of the propagating light is considered to be uniform on the nanometer scale. For cubic QDs, optical selection rules prohibit transitions to states described by even quantum numbers. However, this restriction is relaxed when optical near fields are concerned because of the localized nature of optical near fields in the vicinity of nanoscale matter. Energy in QDs can be optically transferred to neighboring QDs via optical near-field interactions[11]. For instance, assume that two cubic quantum dots—$QD_S$ and $QD_L$, where S and L refer to small and large, and whose side lengths are $a$ and $\sqrt{2}a$, respectively—are located close to each other, as shown in Fig. 1a. Also, suppose that the energy eigenvalues for the quantized exciton energy level specified by quantum numbers $(n_x, n_y, n_z)$ in $QD_S$ are given by

$$E_{(n_x,n_y,n_z)} = E_B + \frac{\hbar^2 \pi^2}{2Ma^2}\left(n_x^2 + n_y^2 + n_z^2\right), \tag{1}$$

where $E_B$ is the energy of the bulk exciton and $M$ is the effective mass of the exciton. A resonance exists between the energy level of $QD_S$ with quantum numbers (1,1,1) (denoted as $S_1$ in Fig. 1a) and that of $QD_L$ with quantum numbers (2,1,1) (denoted as $L_2$ in Fig. 1a). Because of the steep localized electric field in the vicinity of $QD_S$ and $QD_L$, an optical near-field interaction occurs between the two QDs. This interaction is denoted by $U$ in Fig. 1a, and the steep electric field is schematically indicated by the orange triangle. Therefore, energy in $S_1$ can be optically transferred to $L_2$ and vice versa. Normally, such a transition would be dipole forbidden because $L_2$ has an even quantum number. This means that diffraction-limited far-field light irradiation from external systems can couple only to $S_1$[31]. In $QD_L$, optical-energy dissipation, described by $\Gamma$ is faster than the near-field interaction, so the optical energy deposited into the (2,1,1) level can relax to the (1,1,1) level of $QD_L$ (denoted by $L_1$).

Similar optical-excitation transfer via near-field interactions has been reported for various material systems, including CuCl QDs[11], InAs QDs[32], CdSe QDs[33], and hybrid systems[13,14]. Also, the theoretical foundations describing such phenomena, including the optimal near-field interaction that maximizes optical excitation transfer, have been developed by Sangu *et al.* in Ref. [11]. Because the primary focus of the present study is to investigate the possibility of synchronization, chaos, and random-number generation based on optical excitation transfer, we do not assume a particular implementation, as explained in the discussion section. In this study, based on experimental observations of energy transfer in ZnO nanorods[34], we assume a sublevel relaxation $\Gamma^{-1}$ = 10 ps and radiative decay times for $QD_S$ and $QD_L$ of $\gamma_S^{-1}$ = 443 ps and $\gamma_L^{-1}$ = 190 ps, respectively, which are typical values for these parameters. The optical near-field interaction is given by $U^{-1}$ = 120 ps. As shown in Fig. 1b, these parameter values lead to an evolution of populations involving the energy level $L_1$, assuming an initial excitation at $S_1$. These results clearly indicate that optical excitation transfer occurs from $S_1$ to $L_1$.

When the lower level of $QD_L$ ($L_1$) is occupied, the optical excitation in $QD_S$ cannot transfer to that level in $QD_L$ because of the state-filling effect[11]. Optical pulsation based on optical energy transfer forms because of the architecture, where the state filling in $L_1$ is triggered by the radiation from $S_1$ with a delay with respect to the energy transfer from $L_1$, as shown schematically in Fig. 1c. If $QD_S$ is irradiated with cw radiation, such triggers should occur periodically and continuously at constant intervals. In other words, a pulsed signal should result.

We describe the above dynamics by using a density matrix formalism. The radiative relaxation rates from $S_1$ and $L_1$ are denoted as $\gamma_S$ and $\gamma_L$, respectively. The quantum master equation is[35]

$$\frac{d\rho^{NOP}(t)}{dt} = -\frac{i}{\hbar}\left[H_{int} + H_{ext}^{CW}(t) + H_\Delta(t), \rho^{NOP}(t)\right] + \sum_{i=S,L}\frac{\gamma_i}{2}\left[2R_i\rho^{NOP}(t)R_i^\dagger - R_i^\dagger R_i \rho^{NOP}(t) - \rho^{NOP}(t)R_i^\dagger R_i\right]$$
$$+ \frac{\Gamma}{2}\left[2S\rho^{NOP}(t)S^\dagger - S^\dagger S\rho^{NOP}(t) - \rho^{NOP}(t)S^\dagger S\right], \quad (2)$$

where $H_{int}$ represents the interaction Hamiltonian. Matrices $R_i$ ($i$ = S,L) are annihilation operators, which annihilate excitations in $S_1$ and $L_1$, respectively, via radiative relaxations. Matrices $R_i^\dagger$ ($i$ = S,L) are creation operators given by transposes of matrices $R_i$. The matrix $S$ is an annihilation operator that annihilates the excitation in $L_2$ via sublevel relaxations. The external Hamiltonian $H_{ext}^{cw}(t)$ represents the external cw optical excitation that populates the energy level $S_1$ of $QD_S$. This Hamiltonian is given by

$$H_{ext}^{CW}(t) = CW(R_{S_1}^\dagger + R_{S_1}), \quad (3)$$

where CW specifies the amplitude of the external cw radiation. The other external Hamiltonian $H_\Delta(t)$ represents radiation from the lower-energy level of $QD_S$, which affects the lower-energy level of $QD_L$ or $L_2$ with a delay $\Delta$. The Hamiltonian is given by

$$H_\Delta(t) = \alpha\rho_{S_1}^{NOP}(t-\Delta)\left(R_{L_1}^\dagger + R_{L_1}\right), \quad (4)$$

where $\rho_{S_1}^{NOP}(t)$ indicates the population of energy level $S_1$ and $\alpha$ indicates the coupling efficiency. In the original theory of the pulse-generating mechanism[19], the delay line was represented by a different QD combination, giving another density matrix, and the overall dynamics was analyzed by solving the system of equations. In our simplified system, the delay is incorporated into Eq. (4).

In addition to the typical parameter values based on ZnO nanorods[34] introduced earlier, the coupling efficiency α is assumed to be 0.1, and the cw input amplitude is CW = 0.0007. Figure 1d shows the dynamics of the population of the lower level of QD$_L$ (L$_1$) when the interdot optical near-field interaction $U^{-1}$ = 120 ps and Δ = 1000 ps. The population dynamics become pulsed, so we use this model for the following discussion. The period of the oscillating population is approximately 2849 ps, which gives a pulse frequency of approximately 351 MHz.

As reported in Ref. [19], no pulse train occurs for a cw excitation that is either too intense or too weak. Figure 1e shows the peak-to-peak population as a function of the cw input amplitude. Pulses occur for a cw input amplitude between approximately 0.0003 and 0.002.

**Synchronization in Nano-Optics**

We now consider irradiating the NOP with periodic external radiation, as shown schematically in Fig. 2a, and investigate whether synchronization is induced in the system. Consider the system subjected to an external periodic stimulus given by a sinusoidal perturbation:

$$H_T^{\text{Periodic}}(t) = A\sin(2\pi t/T)\left(R_{S_1}^{\dagger} + R_{S_1}\right), \tag{5}$$

where *A* and *T* are the amplitude and period of the periodic signal, respectively. By adding the Hamiltonian represented by Eq. (5) to Eq. (2), synchronization is characterized by solving

$$\begin{aligned}
\frac{d\rho^{\text{Sync}}(t)}{dt} = &-\frac{i}{\hbar}\left[H_{\text{int}} + H_{\text{ext}}^{\text{CW}}(t) + H_{\Delta}(t) + H_{T}^{\text{Periodic}}(t), \rho^{\text{Sync}}(t)\right] \\
&+ \sum_{i=S,L}\frac{\gamma_i}{2}\left[2R_i\rho^{\text{Sync}}(t)R_i^{\dagger} - R_i^{\dagger}R_i\rho^{\text{Sync}}(t) - \rho^{\text{Sync}}(t)R_i^{\dagger}R_i\right] \\
&+ \frac{\Gamma}{2}\left[2S\rho^{\text{Sync}}(t)S^{\dagger} - S^{\dagger}S\rho^{\text{Sync}}(t) - \rho^{\text{Sync}}(t)S^{\dagger}S\right].
\end{aligned} \qquad (6)$$

Let the parameters associated with the NOP be the same as those for the previous discussion. Let the period of the external signal be given by $T$ = 3500 ps (or approximately 286 MHz) and $A$ = 0.0001. The dashed and solid curves in Fig. 2b show the evolution of the population associated with the energy level $L_1$ with and without an external input. The oscillation period is synchronized with the input signal.

While maintaining the parameters associated with the NOP, Fig. 2c characterizes synchronization, or more specifically, the frequency of the external periodic signal that maximizes the spectral peak of the output signal. The dashed curve shows that the oscillating frequency is equal to the frequency of the external signal when the latter is between 244 and 500 MHz. For frequencies outside this locking range, the oscillating frequency is approximately 366 MHz, which is nearly equal to the oscillating frequency of the original NOP exposed to a cw input amplitude of 0.0008. This is consistent with the fact that the system is irradiated with cw radiation (0.0007) in addition to a periodic signal with amplitude 0.0001.

The locking range depends on the amplitude of the external irradiation. The solid, dot-dashed, and dotted curves in Fig. 2c indicate the locking range of synchronization for external irradiation amplitudes $A$ of 0.0015, 0.0012, and 0.0008. The larger the amplitude of the external signal, the larger the bandwidth of synchronization. This property is similar to the mode-locking phenomenon observed in conventional lasers[21] and other systems such as phase-locked loops[36]. Furthermore, we find that an external stimulus with excessively large amplitude does not lead to synchronization; rather, the system is overwhelmed by optical energy and enters a static state.

Figure 2d considers the case in which the amplitude of the external periodic signal is maintained ($A$ = 0.0001). Based on the results shown in Fig. 2d, we investigate the *sensitivity* of the NOP to the external system by changing the amplitude of the original cw pumping light. Recall that the locking range is between 244 and 500 MHz for CW = 0.0007 and an external periodic signal amplitude of 0.0001, respectively. The circles in Fig. 2d show the maximum spectrum obtained when the NOP is exposed to an external input divided by the maximum spectrum of the original pulser without the external input. We refer to this ratio as *sensitivity*, which is larger in the locking range. Moreover, it increases with decreasing cw-excitation power (CW = 0.0006), as indicated by the squares in Fig. 2d. In contrast, as shown by the triangles in Fig. 2d, greater cw excitation power (CW = 0.0008) leads to a decrease in sensitivity. Such properties are also similar to those of conventional mode-locked lasers and are referred to as the *dependence on relative excitation*[37].

These results clearly imply that the physics of near-field optical systems can lead to synchronized phenomena.

**Chaos in Nano-Optics**

Lasers are known to undergo chaotic oscillation when connected with a delayed feedback[21,23]. Here, we address the question of whether chaos is possible in the subwavelength regime. In other words, we investigate the possibility of chaos evolving from nanoscale optical-energy transfer.

When an external delay line is added to the original NOP system, as shown schematically in Fig. 3a, the overall dynamics are described by solving

$$\begin{aligned}
\frac{d\rho^{\text{Chaos}}(t)}{dt} = &-\frac{i}{\hbar}\left[H_{\text{int}} + H_{\text{ext}}^{\text{CW}}(t) + H_{\Delta}(t) + H_{\Delta_C}(t), \rho^{\text{Chaos}}(t)\right] \\
&+ \sum_{i=S,L}\frac{\gamma_i}{2}\left[2R_i\rho^{\text{Chaos}}(t)R_i^{\dagger} - R_i^{\dagger}R_i\rho^{\text{Chaos}}(t) - \rho^{\text{Chaos}}(t)R_i^{\dagger}R_i\right] \\
&+ \frac{\Gamma}{2}\left[2S\rho^{\text{Chaos}}(t)S^{\dagger} - S^{\dagger}S\rho^{\text{Chaos}}(t) - \rho^{\text{Chaos}}(t)S^{\dagger}S\right],
\end{aligned} \quad (7)$$

where the lower energy level L$_2$ of QD$_L$ is fed back to the same energy level after time delay $\Delta$. This effect is taken into account by adding the following external Hamiltonian to the original master equation:

$$H_{\Delta_C}(t) = -\alpha_C \rho_{L_1}^{\text{Chaos}}(t - \Delta_C)\left(R_{L_1}^{\dagger} + R_{L_1}\right), \quad (8)$$

where $\alpha_C$ and $\Delta_C$ are the coupling constant and timing delay, respectively. The quantity $\rho_{L_1}^{\text{Chaos}}(t)$ is the population of L$_1$.

Parameter values for the systems are based on experimental observations from ZnO nanorods[28]: $\gamma_S^{-1}$ = 443 ps, $\gamma_L^{-1}$ = 190 ps, $\Gamma^{-1} = 10$ ps, $\Delta$ = 3000 ps, $\alpha$ = 0.1, and $U^{-1}$ = 100 ps. Figure 3b considers the situation in which delay lines with $\Delta_C$ = 1000 ps are incorporated. Figure 3b shows populations when the coupling constant $\alpha_C$ is 0.001, 0.01, 0.02, and 0.05 (see Fig. 3b-i, 3b-ii, 3b-iii, and 3b-iv, respectively). Case (i) exhibits a periodic signal, whereas case (iv) converges to a constant population. Cases (ii) and (iii) exhibit more complicated dynamics.

To quantitatively characterize the dynamics, we evaluate the local maxima and minima of populations as a function of the coupling constant $\alpha_C$. When the population dynamics is periodic or constant, there is no diversity in the local maxima and minima, whereas the maxima and minima take on a variety of values when the signal is chaotic[21], which leads to bifurcations and chaos in signal trains.

The circles and crosses in Fig. 3c show the local maxima and minima, respectively, in the population between 500 000 and 1 000 001 ps. For $\alpha_C$ between 0.001 and approximately 0.0025, the variation in local maxima and minima is limited, whereas for $\alpha_C$ = 0.0025, the variation is greater. From $\alpha_C$ > ~0.0025 to nearly 0.009, the variation is again limited, whereas from approximately 0.009 to 0.0228, the variation increases again. Beyond $\alpha_C$ = 0.0228, the local maxima and minima have similar values, so no oscillations occur. These results indicate that a system based on optical energy transfer exhibits bifurcation and chaotic behavior, which is evidence of chaos.

Another criterion satisfied by chaos is expressed by the maximal Lyapunov exponent (MLE)[22,23]. Suppose that a trajectory exhibits chaotic behavior, which means that the final difference between two trajectories with a subtle initial difference $\delta \boldsymbol{Z}_0$ grows exponentially. In other words, $|\delta \boldsymbol{Z}(t)| \approx \exp(\lambda t)|\delta \boldsymbol{Z}_0|$. The MLE is defined by

$$\lambda = \lim_{t \to \infty} \lim_{\delta \boldsymbol{Z}_0 \to \infty} \frac{1}{t} \ln \frac{|\delta \boldsymbol{Z}(t)|}{|\delta \boldsymbol{Z}_0|}, \qquad (9)$$

where $\lambda \leq 0$ indicates no chaos[20,22]. We used the FET1 code developed by Wolf *et al.*[38] to estimate the MLE from a time series. Figure 4a shows the calculated MLE as a function of the control parameter $\alpha_C$. The results show that, for instance, $\lambda$ is positive for 0.0148 < $\alpha_C$ < 0.0225. This particular range coincides with the range over which chaos occurs in the local maxima and minima (Fig. 3b). Also, for the 78 points in this particular regime, there are 26 points that satisfy the random-number conditions discussed below. This result clearly indicates that the physics of near-field optics allows for chaotic phenomena.

**Random-Number Generation by Nano-Optics**

Finally, to determine if NOPs can be used as RNGs, we use statistical-test suites to evaluate the randomness inherent in the chaotic dynamics of populations. Many well-known statistical-test suites, such as NIST 800-22[39,40], FIPS 140-2[41,42], and Diehard[43], are proposed in the literature. We use the FIPS 140-2 statistical-test suite (hereafter the "FIPS test") because it is the most simple and user-friendly test suite. It offers the following four basic tests: (i) *the monobit test*, (ii) *porker test*, (iii) *run test*, and (iv) *long-run test*. Because of its simplicity, it has been used to supplement RNGs in many hardware implementations[44,45].

Kim *et al.* recalculated the requirement of the FIPS test for a 2500-bit sequence to give an identical significance level $\alpha = 10^{-2}$, which is a commonly used value in cryptography. This significance level is defined as the probability of a false rejection of the null hypothesis in a statistical test. In other words, it is the probability that a perfect RNG generates a "failure" sequence. A summary of the "improved FIPS test," which we use in the analysis, is available in the supplementary information, and further details may be found in Refs. [46] and [47].

We obtained temporal signals from $t = 0$ to $t = 1\,000\,000$ for each member of the population with a resolution of 1 ps. The population at each time increment is converted to a 16-bit-precision fixed-point number and the lowest significant bit is used for a binary value. In other words, $1\,000\,001$ binary bits are obtained from a single run. Ignoring the initial time range from $t = 0$ to $100\,000$, the total length is reduced to $900\,000$ bits. The signals are then divided into increments of 2500 (i.e., 2500 ps duration); the number of 2500-bit binary-signal sets is 36. By subjecting all 36 sets to the improved FIPS test described above, we can determine whether they qualify as random numbers. If all bit sets pass the test or if the number of failures is at most two for each test, then the answer is yes. Two failures are deemed acceptable in this particular case because the acceptable interval is determined to be in the 99.73% range of normal distribution. For details, see section 4.2.1 of Ref. [39].

The solid, dashed, dotted, and dot-dashed curves in Fig. 4b show the frequency of the FIPS-test failure for the monobit, poker, run, and long-run tests, respectively, as a function of the control parameter $\alpha_C$. The FIPS test is passed in a total of 35 cases, which used the following control parameters: 0.0058, 0.0061, 0.0076, 0.0084, 0.0104, 0.0106, 0.0108, 0.0116, 0.0124, 0.0172, 0.0175, 0.0177–0.0180, 0.0181, 0.0183, 0.0185, 0.0187, 0.0190, 0.0192, 0.0194–0.0196, 0.0198–0.0200, 0.0202, 0.0204–0.0207, and 0.0212–0.0214. The evaluation was based on an $\alpha_C$ interval of $10^{-3}$, as shown in Fig. 4c. Moreover, for all control parameters for which the improved FIPS test was passed, the corresponding MLEs are positive.

The chaotic signal behaves differently depending on other parameters. Focusing on the external delay, which plays a crucial role in generating chaos, Fig. 5 characterizes the pass–no-pass results of the FIPS test for time delay $\Delta_C$ ranging from 0 to 3000 ps. The number of cases that pass the FIPS test is given on the right side of Fig. 5 (where $\Delta_C$ = 200 ps yields the maximum number of cases that pass the FIPS test). From this analysis, we conclude that chaotic phenomena based on near-field local optical interactions can form the basis of ultrasmall RNGs.

**Discussion**

As mentioned in the introduction, complex oscillatory dynamics are observed in various systems in the real world. Our study indicates that local nanoscale interactions may lead to synchronization, chaos, and even random-number generation. The optical near-field interactions examined in this study contain the "necessary conditions" required for generating complex oscillatory dynamics; that is, nanoscale optical near-field interactions generate physical properties that are functionally equivalent to those observed in other physical systems exhibiting complex dynamics. Moreover, using the optical near field yields the additional benefit of overcoming the diffraction limit of light.

However, several important unresolved issues remain that complicate the science of near-field optics. The function of delay, in particular, needs more study, both theoretically and experimentally. Optical excitation transfer depends on the architecture of nanostructures as well as on the population of energy levels therein. For example, in Ref. [48], Naruse *et al.* discusses topology-dependent, autonomous optical excitation transfer, and how an excitation "waits" in a multi-quantum-dot system. In addition, the coupling between semiconductor quantum dots and nanocavities has been intensively studied[49,50,51]. These investigations are analogous to the present work in that they consider how to realize further functionalities by using near-field optics, but we need to be careful because the notion of "cavity" implies diffraction-limited macroscale entities. We will thus conduct further investigations into the theoretical foundation of delay in near-field optics. Experimentally, on the other hand, Nomura *et al.* demonstrated a chain of colloidal CdSe QDs[52]. Also, QD arrangements of graded size have been demonstrated by Kawazoe *et al*.[53] and Franzl *et al*.[54], which could provide the basic resources to implement delay functions[19]. To fabricate devices in the future, fluctuations in size, layout, etc., in the quantum nanostructure may be of concern, and tolerance and robustness would need to be clarified. A step in this direction has already been taken by Naruse *et al.*, who built a stochastic model to systematically characterize optical excitation transfer in multilayer InAs QDs formed by molecular beam epitaxy[55].

Other unique optical near-field processes can be considered. For example, the hierarchical properties of the optical near field means that near-field interactions behave differently depending on the length scale involved[56]. This property is notably different from that encountered in conventional optics and photonics. Another interesting topic is the impact of the hierarchical properties of optical near fields on synchronization and chaos. As techniques to fabricate nanophotonic devices continue to be developed, experimental verification and fabrication of practical devices are important routes for future work[13–16,29,30,34,55].

**Acknowledgment**

This work was supported in part by the Core-to-Core Program, A. Advanced Research Networks from the Japan Society for the Promotion of Science.

**Author contributions**

M.N. and S.-J.K directed the project; M.N. and S.-J.K. designed systems; M.N. and S.-J.K. analyzed the data; M.A. H.H. and M.O. discussed physical modeling; M.N and S.-J.K wrote the paper.

**Competing financial interests**

The authors declare no competing financial interests.

**Figure Captions**

**Figure 1 | Nanoscale optical pulser architecture.** (a) Optical excitation transfer via near-field interactions between closely located smaller and larger quantum dots (QDs). (b) Example of optical excitation transfer from a smaller to a larger QD. (c) By incorporating a time delay, optical pulsation becomes possible. (d) Example of optical pulses induced by cw optical excitation. (e) Peak-to-peak value of pulsed population as a function of cw excitation amplitude.

**Figure 2 | Synchronization in NOP.** (a) Schematic of system where NOP is subjected to external periodic signal. (b) Evolution of population with and without external input. (c) Synchronization of NOP to external input radiation. The bandwidth of the frequency locking increases with the amplitude of the external input. (d) Analysis of sensitivity of synchronization. Synchronization of weakly excited NOP is more sensitive to external input.

**Figure 3 | Chaos in nano optical pulser.** (a) Schematic of system where NOP is connected with external delay. (b) Evolution of population with four different parameters: (i) Periodic signal occurs. (ii) and (iii) Rather complex trains occur. (iv) Population is saturated at a certain level. (c) Local maxima and minima of populations as a function of control parameter $\alpha_C$.

**Figure 4 | Chaos and random-number generation in nanosized system.** (a) Lyapunov exponents as a function of control parameter $\alpha_C$. We used the following FET1 parameters, dimension = 7, delay = 10, evolve = 1, Scale$_{min}$ = $10^{-5}$, and Scale$_{max}$ = 0.7. The Lyapunov exponent $\lambda \leq 0$ indicates no chaos. The dotted line shows $\lambda = 0$. (b) Analysis of properties of random numbers based on the improved FIPS test. (c) Schematic of cases that pass the improved FIPS test. For all 35 cases

that pass the improved FIPS test, the corresponding Lyapunov exponent is positive [see panel (a)].

**Figure 5 | Distributions of cases that pass the improved FIPS test.** The randomness observed in output populations depends on the external delay $\Delta_C$ and control parameter $\alpha_C$. The squares at the right give the number of case that pass the improved FIPS test as a function of external delay.

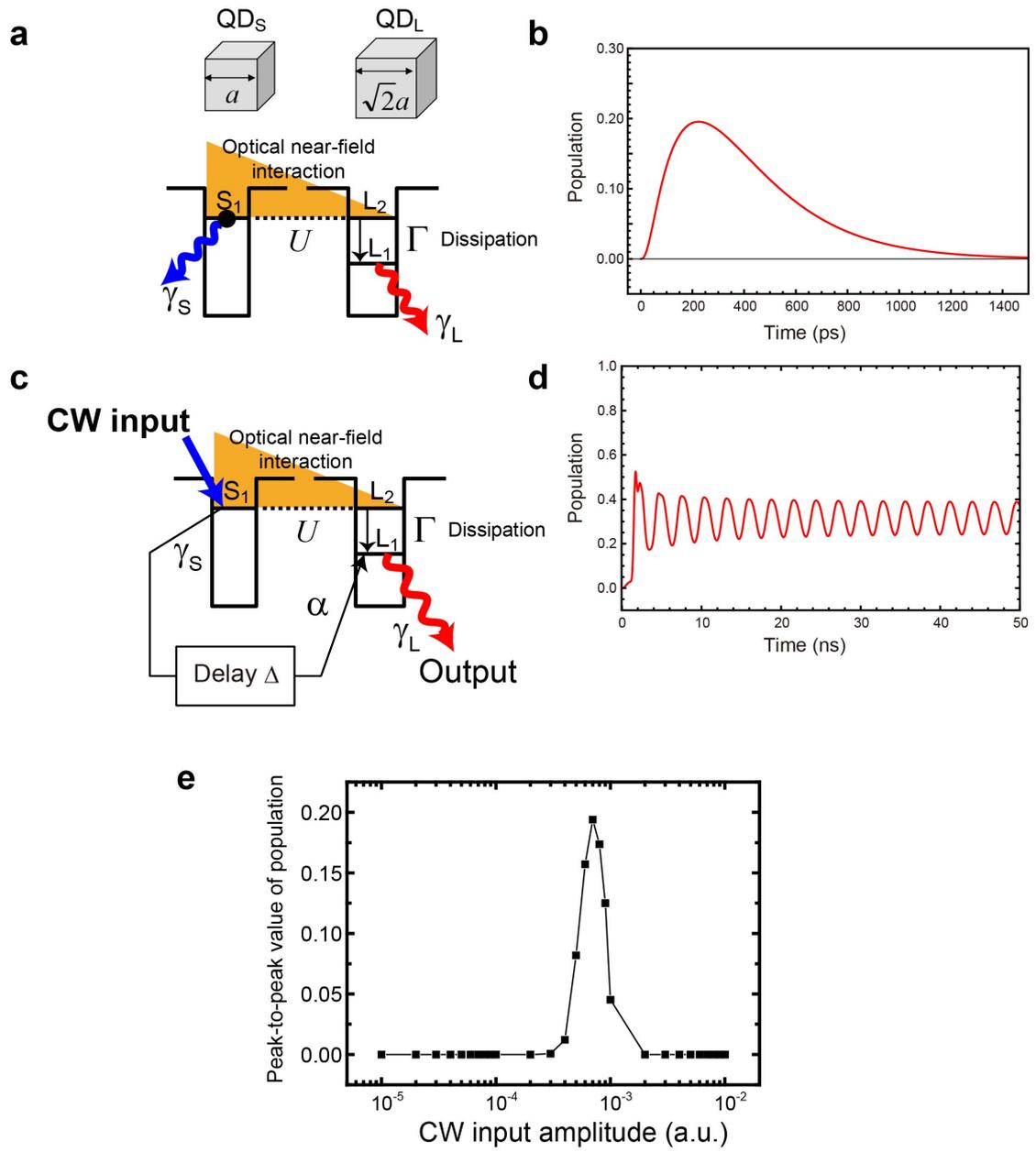

Figure 1

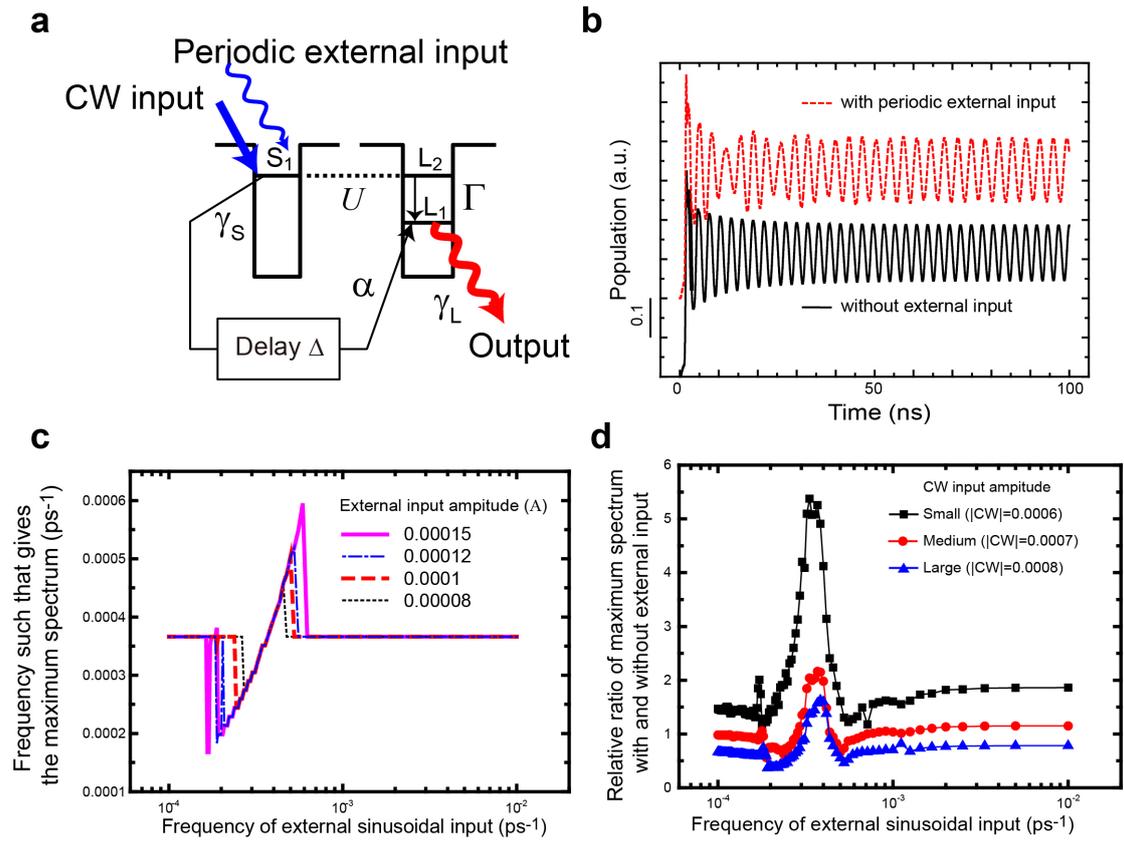

Figure 2

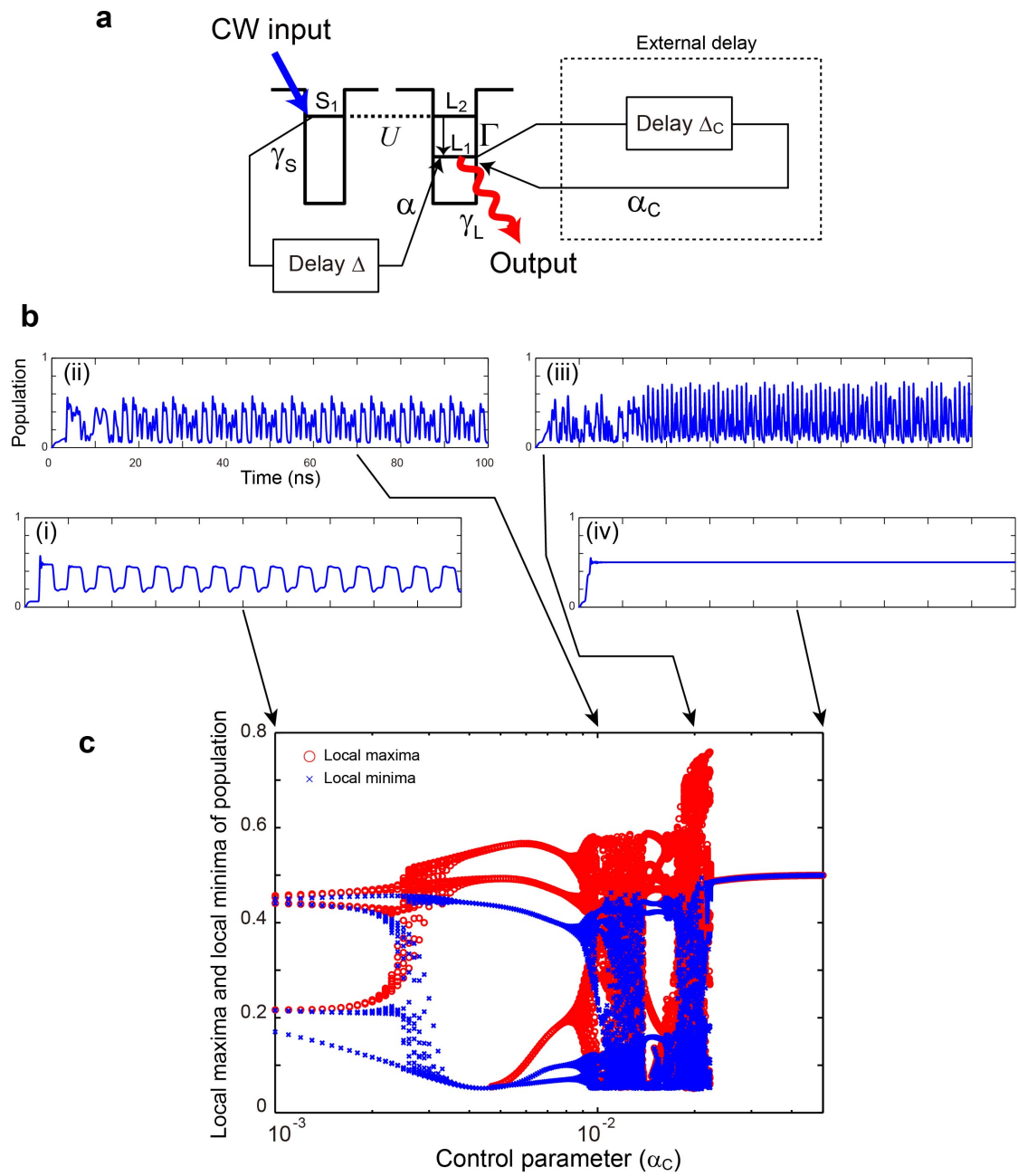

Figure 3

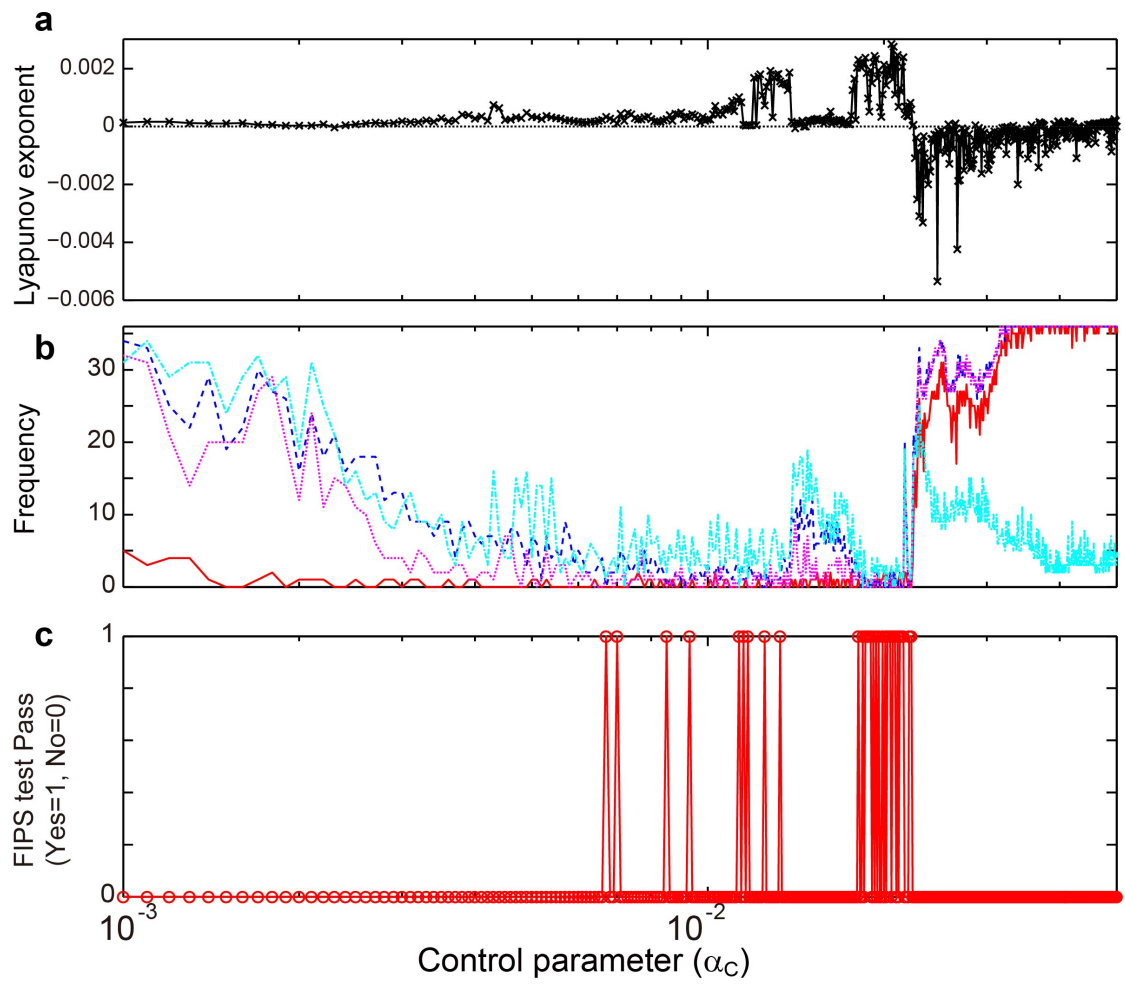

Figure 4

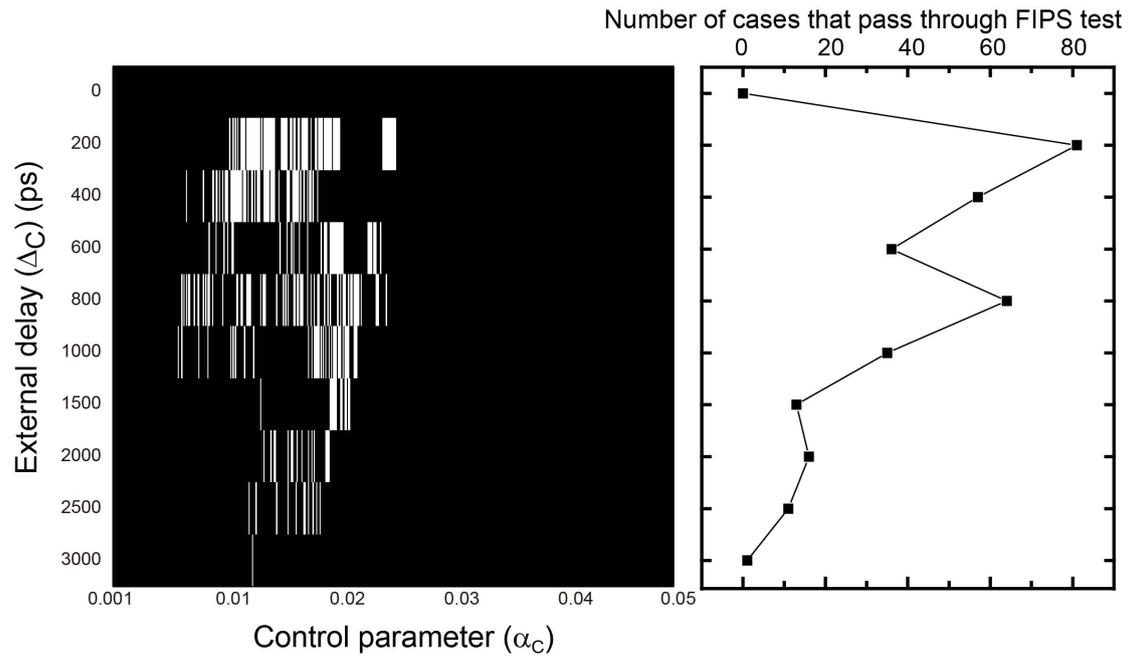

Figure 5